\documentclass[11pt]{article}
\usepackage{amssymb}
\begin{document}
\def\nmonth{\ifcase\month\ \or January\or
   February\or March\or
April\or May\or June\or July\or August\or
   September\or October\or
November\else December\fi}
\def\nmonth{\ifcase\month\ \or January\or
   February\or March\or April\or May\or June\or July\or August\or
   September\or October\or November\else December\fi}
\def\rightheadline{\hfill\folio\hfill}
\def\leftheadline{\hfill\folio\hfill}
\newtheorem{theorem}{Theorem}[section]
\newtheorem{lemma}[theorem]{Lemma}
\newtheorem{remark}[theorem]{Remark}
\def\operatorname#1{{\rm#1\,}}
\def\text#1{{\hbox{#1}}}
\def\qedbox{\hbox{$\rlap{$\sqcap$}\sqcup$}}
\def\BB{{\mathcal{B}_1}}
\def\CC{{\mathcal{B}_2}}
\def\BX{{\mathcal{B}_{DR}}}
\def\BD{{\mathcal{B}_D}}
\def\BR{{\mathcal{B}_R}}
\def\B{{\mathcal{B}}}
\def\tr{{\operatorname{Tr}}}
\def\dvol{{\operatorname{dvol}}}
\newcommand{\reals}{\mathbf{R}}
\newcommand{\nats}{\mbox{${\rm I\!N }$}}

\def\gat{\gamma_a^T}
\def\la{\lambda}
\def\om{\omega}
\def\La{\Lambda}
\def\Th{\theta}
\def\chs{\chi^\star}
\def\dirac{D\!\!\!\!/}
\def\pip{\Pi_+}
\def\pim{\Pi_-}
\def\pipl{\Pi_+^\star}
\def\pimi{\Pi_-^\star}
\def\gf{\tilde{\gamma}}
\def\rand{\left|_{\partial M}\right. }
\def\ch{\cosh \Theta}
\def\sh{\sinh\Theta}
\def\nen{\Lambda \ch -2\om\sh}
\def\tr{\mbox{Tr}}

\newcommand{\gi}{\gamma_i}
\newcommand{\ga}{\gamma_a}
\newcommand{\gt}{\tilde\gamma}
\newcommand{\gm}{\gamma_m}
\newcommand{\beq}{\begin{eqnarray}}
\newcommand{\eeq}{\end{eqnarray}}
\newcommand{\nn}{\nonumber}

\makeatletter
  \renewcommand{\theequation}{%
   \thesection.\arabic{equation}}
  \@addtoreset{equation}{section}
 \makeatother

\title{Strong ellipticity and spectral properties of chiral bag boundary
conditions}
\author{C.G. Beneventano\thanks{E-mail: gabriela@obelix.fisica.unlp.edu.ar}\\
Departamento de F\'{\i}sica, Universidad Nacional de La Plata\\
C.C.67, 1900 La Plata, Argentina
$$\\[.3cm]
P.B. Gilkey\thanks{
E-mail: gilkey@darkwing.uoregon.edu}\\
 Department of Mathematics, University of
Oregon\\
Eugene OR 97403 USA
$$\\[.3cm]
K. Kirsten \thanks{
E-mail: Klaus\_Kirsten@Baylor.edu}\\
Department of Mathematics, Baylor University \\
Waco, TX 76798, USA\\[.3cm]
E.M. Santangelo\thanks{E-mail: mariel@obelix.fisica.unlp.edu.ar}\\
Departamento de F\'{\i}sica, Universidad Nacional de La Plata\\
C.C.67, 1900 La Plata, Argentina
}
\maketitle
\begin{abstract}
We prove strong ellipticity of chiral bag boundary conditions on
even dimensional manifolds. From a knowledge of the heat kernel in
an infinite cylinder, some basic properties of the zeta function
are analyzed on cylindrical product manifolds
of arbitrary even dimension.\\[.3cm]
{\bf Subject Classification} 
\end{abstract}
\section{Introduction}
The influence that boundary conditions have on different spectral
functions is an active field of research. In quantum field theory,
spectral functions of particular interest are the zeta function
and the heat kernel. Their dependence on the boundary condition is
well understood for a large variety of boundary conditions (for a
recent review see \cite{kirs01}). However, for some boundary
conditions an understanding of elementary properties of spectral
functions is still lacking. This is the case for generalized (or
chiral) local bag boundary conditions
\cite{hras84-245-118,wipf95-443-201}.

These boundary conditions involve an angle $\theta$, which is a
substitute for introducing small quark masses to drive the
breaking of chiral symmetry
\cite{wipf95-443-201,duer97-255-333,duer99-273-1}. The influence
the parameter $\theta$ has on various correlators was analyzed in
detail in \cite{wipf95-443-201} for the two dimensional Euclidean
ball. In reference \cite{bene02-35-9343}, the heat kernel and the
eta function were analyzed in the two dimensional cylinder. The
situation of an arbitrary dimensional ball was considered in
\cite{espo02-66-085014}. The results found in the above articles
suggest that general properties of spectral functions, like the
pole structure of the associated zeta function and the form of the
asymptotic expansion of the heat kernel, are the properties
resulting from strongly elliptic local boundary conditions. For
the generalized bag boundary conditions considered here, this
property has not been proven. In fact, strong ellipticity is not
clear at all because, for $\theta \neq 0$, the boundary conditions
are of mixed oblique type \cite{espo02-66-085014} and, under
certain circumstances, oblique boundary conditions are not
strongly elliptic \cite{dowk99-16-1917,avra99-200-495}. However,
after introducing some basic notation and properties of the
boundary conditions, we will prove in Sections 3 and 4 that
generalized bag boundary conditions are indeed strongly elliptic
boundary conditions. Based on this observation, in future
investigations one might envisage a determination of heat kernel
coefficients for these boundary conditions. In recent years, a
conglomerate of methods has been proven to be very effective in
the determination of this asymptotics \cite{kirs01}. Among the
methods are special case considerations, which form the basis of
the second half of our paper. In particular, we will determine the
local heat kernel and the zeta function on cylindrical product
manifolds. Results are given in terms of boundary data, much in
the way it is possible for spectral boundary conditions
\cite{grub96-6-31}. The Conclusions will summarize the main
results and describe their possible future applications. We refer
to \nocite{1,2a,2b,3,5,6,7,8,9,12} [11--20] for other important
related work in the field.

\section{Basic properties of chiral bag boundary conditions}\label{Sect1}

In this section, we will establish some notations for the problem
at hand, i.e., the Euclidean Dirac operator acting on spinors
satisfying local (chiral bag) boundary conditions
\cite{wipf95-443-201}.

Let $m=2\bar m$ be even and let $P =i\gamma_j \nabla_j $ be an
operator of Dirac type on a compact oriented Riemannian manifold
of dimension $m$. Let $V$ denote the spinor space; $dim (V) =
2^{\bar m}$.  An explicit representation of the $\gamma$-matrices
is provided in Appendix A. They are self-adjoint and satisfy the
Clifford anti-commutation relation (\ref{A3}). Near the boundary,
let $e_m$ be the inward unit normal and let $\gm$ be the
projection of the $\gamma$-matrix on the inward unit normal. In
addition let $\tilde\gamma$ be the generalization of `$\gamma_5$'
to arbitrary even dimension, $\tilde\gamma = (-i)^{\bar m}
\gamma_1 ... \gamma_m$.

We set \beq \chi = i \gt e^{\theta \gt} \gm \nn \eeq and use the
relation $\gi \gt + \gt \gi =0$ to compute \beq \chi ^2 = - \gt
e^{\theta \gt} \gm \gt e^{\theta \gt} \gm = \gt e^{\theta \gt} \gt
e^{-\theta \gt} \gm \gm = {\bf 1} .\nn \eeq We define
$$\Pi _\pm := \frac 1 2 ( 1\pm \chi ) $$
and have
$$\Pi_\pm ^2 = \Pi _\pm \mbox{ and } \Pi_- \Pi _+ = \Pi _+ \Pi _- = 0.$$

Note that these two projectors are not self-adjoint (except for
the particular case $\Th =0$). Rather, calling their respective
adjoints $\pimi$ and $\pipl$, one has $\Pi _\pm ^{\star}:= \frac 1
2 ( 1\pm i \gt e^{-\theta \gt} \gm )$, and the following equations
hold

\beq\begin{array}{l} \pipl \pip=\cosh{(\Th \gf)}\exp{(-\Th \gf)}
\pip=\pipl \cosh{(\Th \gf)}\exp{(-\Th \gf)}\nn\\\pimi
\pip=\sinh{(\Th \gf)}\exp{(-\Th \gf)} \pip=\pimi \sinh{(\Th
\gf)}\exp{(-\Th \gf)}\nn\\\pimi \pim=\cosh{(\Th \gf)}\exp{(-\Th
\gf)} \pim=\pimi \cosh{(\Th \gf)}\exp{(-\Th \gf)} \nn\\\pipl
\pim=\sinh{(\Th \gf)}\exp{(-\Th \gf)} \pim=\pipl \sinh{(\Th
\gf)}\exp{(-\Th \gf)}\,.\end{array}\label{prod}\eeq

\bigskip

 We use $\Pi _-$ to define boundary conditions for $P$.
Similarly, we shall let ${\cal B} := \Pi _- \oplus \Pi _- P $
define the associated boundary condition for $P^2$.

In the following two sections, we will show that $(P, \Pi _-)$ and
$(P^2 , {\cal B})$ define strongly elliptic boundary conditions
and, as a result, we can assume standard results on the
meromorphic structure of eta and zeta-invariants hold
\cite{gilk95b}. Otherwise stated, we will prove statements (1) and
(2) of the following Theorem
\newtheorem{sell}{Theorem}[section]
\begin{sell}
\label{sell}
$\phantom{a}$\\
(1) $(P,\Pi _-)$ is strongly elliptic with respect to $\mathbf{C}-\mathbf{R}^+ - \mathbf{R}^-$.\\
(2) $(P^2 , {\cal B})$ is strongly elliptic with respect to $\mathbf{C}-\mathbf{R}^+$.\\
(3) $(P, \Pi _-)$ is self-adjoint.\\
(4) $(P^2 , {\cal B})$ is self-adjoint.
\end{sell}
Statements (3) and (4) are well known to hold
\cite{wipf95-443-201} and so we concentrate on statements (1) and
(2).

\section{Ellipticity of the first order boundary value problem}
\label{Sect2}

{\bf Proof of (1):} We use Lemma 1.11.2 (a) of \cite{gilk95b} to
prove assertion (1). Note that the special case $\theta =0$
defines standard mixed boundary conditions and the theorem is
known to hold for this case. Let $x=(y_1,...,y_{m-1},x_m)$ be
coordinates near the boundary where $x_m$ is the geodesic distance
to the boundary and where $y=(y_1,...,y_{m-1})$ are coordinates on
$\partial M$. Let $\xi_a dy^a \in T^* (\partial M)$.

Following \cite{gilk95b} we define \beq \tilde q (\xi ,\lambda ) =
-i\gm \left(\sum_{a<m} \ga\xi_a -\lambda \right) \mbox{ for
}(\xi,\lambda)\neq (\vec 0 , 0) \mbox{ and } \lambda\not\in
\mathbf{R} - \{0\}.\nn \eeq We then have $\tilde q (\xi , \lambda
)^2 = (|\xi | ^2 -\lambda ^2 ) {\bf 1}$. As $(|\xi | ^2 -\lambda
^2 ) \not\in i\mathbf{R}$, we may let $V_\pm^{\tilde q}$ be the
span of the eigenvectors of $\tilde q (\xi ,\lambda )$ with
positive/negative real parts. We let \beq W:= \mbox{Kernel} (\Pi
_- ) = \mbox{Range} (\Pi _+ ) .\nn \eeq Using Lemma 1.11.2 (a) of
\cite{gilk95b}, we prove assertion (1) by verifying that $\Pi _-$
is an isomorphism from $V_-^{\tilde q} (\xi , \lambda ) $ to
${\cal W} = \mbox{Range} (\Pi _- )$. This is equivalent to showing
\beq V_- ^{\tilde q} (\xi , \lambda ) \cap W = \{0\}
.\label{assert1} \eeq We change variables slightly setting
$\lambda = -i\mu$ where $\mu \not\in i\mathbf{R} - \{0\}$ and
$(\xi , \mu ) \neq (\vec 0 , 0)$. We then have \beq \tilde q (\xi
, \mu ) = -i\gm \sum_{a<m} \ga \xi_a + \mu \gm \mbox{ and } \tilde
q (\xi , \mu )^2 = (|\xi | ^2 + \mu ^2 ) {\bf 1} .\nn \eeq The
next step in the proof is to reduce the problem to a collection of
effective two-dimensional ones. We make use of the properties of
the $\gamma$-matrices. First note that the elements \beq \tau _1
:= i \gamma_2 \gamma_3, \quad ..., \tau_{\bar m -1} :=
i\gamma_{m-2} \gamma_{m-1}, \nn \eeq mutually commute and, in
addition, they commute with $\gamma_1$ and $\gm$. Thus, for
$j,k=1,...,\bar m -1$, we have \beq
\tau _j \tau _k = \tau _k \tau _j , & \quad & \tau _j ^2 =1 , \nn\\
\tau_j \gamma_1 = \gamma_1 \tau _j , & \quad & \tau_j \gm = \gm
\tau_j .\nn \eeq So we can choose a set of simultaneous
eigenvectors of $\tau_j$ with eigenvalues $\rho_j = \pm 1$. We
denote by $\vec \rho = (\rho_1 , ..., \rho_{\bar m -1})$ the
collection of simultaneous eigenvalues of $\tau_j$ and we define
the associated simultaneous eigenspaces by \beq V_{\vec \rho} = \{
v \in V: \tau_i v = \rho _i v\} .\nn \eeq The vector space
$V_{\vec \rho}$ is preserved by $\gamma_1$, $\gm$ and $\gt$. We
use this fact to decompose $V_{\vec \rho}$ into its chiral parts,
\beq V_{\vec \rho} = V_{\vec \rho}^+ \oplus V_{\vec \rho} ^-
\mbox{ where }V_{\vec \rho}^\pm = \{ v\in V_{\vec \rho}: \gt v =
\pm v\}. \nn \eeq Let $V_{\vec \rho}^+ = \mbox{span} \{v_{\vec
\rho}\} $. Then since $\gm \gt = -\gt \gm$ we have \beq V_{\vec
\rho} = \mbox{span} \{ v_{\vec \rho} , \gm v_{\vec \rho}\}.\nn
\eeq The vector spaces $V_{\vec \rho}$ provide the decomposition
\beq V= \bigoplus_{\vec \rho}V_{\vec \rho} , \quad \mbox{dim}
V_{\vec \rho} =2 ,\nn \eeq and the problem completely decouples
into two-dimensional spaces.

On $V_{\vec \rho}$ one easily computes, using the definition
$\epsilon (\vec \rho )= \rho_1\times...\times\rho_{\bar m -1}$,
\beq
\gamma_1 = \epsilon (\vec \rho ) \left(
\begin{array}{cc}
0 & i \\
-i & 0
\end{array}\right), \quad
\gm = \left( \begin{array}{cc}
0 & 1 \\
1 & 0
\end{array}\right) ,\quad
\gt = \left( \begin{array}{cc}
1 & 0 \\
0 & -1
\end{array}\right),\nn
\eeq thus reproducing the two-dimensional Pauli-matrices up to the
standard sign ambiguity.

We note that the kernel of $\Pi _-$ is determined by the
eigenvectors of $\chi$. Thus to establish (\ref{assert1}) we shall
need explicit representations of $\chi$ and $\tilde q (\xi , \mu)$
acting on $V_{\vec \rho}$. To calculate $\tilde q (\xi , \mu)$ it
is possible to choose coordinates such that $\xi _2 = ... = \xi
_{m-1}=0$. It is then easy to see that \beq \chi = i \left(
\begin{array}{cc}
0 & e^\theta \\
-e^{-\theta} & 0
\end{array}
\right) , \quad
\tilde q (\xi , \mu ) =
\left( \begin{array}{cc}
\epsilon (\rho) \xi _1 & \mu \\
\mu & -\epsilon (\rho ) \xi _1
\end{array} \right) .\nn
\eeq The eigenvectors of $\chi$ follow, \beq \chi {\varrho \choose
-i e^{-\theta} } = \varrho {\varrho \choose -i e^{-\theta} } , \nn
\eeq and we compute \beq \tilde q (\xi , \mu ) {\varrho \choose -i
e^{-\theta} } = { \epsilon (\vec \rho ) \varrho \xi_1 - i \mu
e^{-\theta} \choose \mu \varrho + i\epsilon (\vec \rho ) \xi _1
e^{-\theta} }\nn \eeq where $\varrho = \pm 1$. Assertion (1) holds
if these are not multiples of each other, that is if \beq \det
\left(
\begin{array}{cc}
 \epsilon (\vec \rho ) \varrho \xi_1 - i \mu e^{-\theta} & \varrho \\
\mu \varrho + i\epsilon (\vec \rho ) \xi _1 e^{-\theta} & -i e^{-\theta }
\end{array}
\right) = -2i\epsilon (\vec \rho ) \varrho \xi _1 e^{-\theta} -
\mu \left(1+e^{-2\theta } \right) \neq 0 .\nn \eeq Since the first
term is purely imaginary and $\mu \not\in i\mathbf{R} - \{0\}$
this proves assertion (1).

In fact, to prove what we announced (i.e., that $\Pi _-$ is an
isomorphism from $V_-^{\tilde q} (\xi , \lambda ) $ to ${\cal W} =
\mbox{Range} (\Pi _- )$), it is enough to consider only $
\varrho=+1$. But since $V_-^{\tilde q} (-\xi , -\lambda
)=V_+^{\tilde q} (\xi ,\lambda )$, considering $\rho = \pm 1$
gives the same condition.

Although assertion (2) follows from assertion (1) and Lemma 1.11.2
(b) in \cite{gilk95b}, we prefer to give a second proof showing
that the boundary operator ${\cal B}$ involves tangential
derivatives. This makes apparent that the chiral boundary
conditions are non-standard boundary conditions.

\section{Ellipticity of the second order boundary problem}
\label{Sect3}

When considering spectral properties of the square of the operator
of  Dirac type, $P^2= (i\gamma_j\nabla_j )^2$, the boundary
condition imposed through ${\cal B}$ is \beq
\pim \psi \rand &=& 0 ,\label{bc1}\\
\pim \gamma_j\nabla_j \psi \rand &=& 0 . \label{bc2} \eeq The
second boundary condition (\ref{bc2}) can be rewritten as an
oblique boundary condition involving tangential derivatives. To
this end note \beq
\gm \Pi_\mp &=& \Pi_\pm ^\star \gm , \nn\\
\ga\Pi_\mp &=& \Pi_\mp^\star \ga , \nn \eeq with the tangential
$\gamma$-matrices $\ga$, where $a=1,...,m-1$. This allows the
boundary condition (\ref{bc2}) to be written as \beq 0&=& \pim
(-i\gamma_j \nabla_j) \psi \rand = i\gm \pim (-i \gamma_j
\nabla_j)
\psi \rand \nn\\
&=& \gm \pim (\gm\nabla_m +\ga \nabla_a ) \psi \rand = (\pipl
\nabla_m
+\gm \ga \pimi \nabla_a ) \psi \rand \nn\\
&=&(\pipl \nabla_m +\gm \ga \pimi \nabla_a ) (\Pi_- + \pip ) \psi
\rand ,\label{bc3} \eeq where $\nabla_m$ is the interior normal
derivative. The boundary condition contains tangential derivatives
and the conditions imposed through ${\cal B}$ could thus be termed
of mixed oblique type.

{\bf Proof of (2):} To study the ellipticity of the boundary value
problem, we introduce the ``partial" leading symbol of $P^2$ \beq
\sigma_L (y,x_{m},\omega, -i \partial_m, \lambda ) = -\partial_m^2
+\omega^2 -\lambda \nn \eeq and the graded symbol, $\sigma_g$, of
${\cal B}$ \beq \sigma_g = \left(\begin{array}{cc}
  \pim& 0 \\
  -\ga \omega_a  \pimi& i \gm  \pipl
\end{array}\right)\,.\nn
\eeq Strong ellipticity requires that the problem \beq \sigma_L
(y,x_m,\omega, -i \partial_m, \lambda
)\Psi(y,x_m,\omega,\lambda)=0 \label{ed}\eeq with \beq
\Psi\to_{r\to \infty} 0 \label{ci}\eeq and \beq \left.\sigma_g
\left(\begin{array}{c}
  \Psi \\
  \partial_m \Psi
\end{array}\right)\right\rfloor_{r=0}=\left.\left(\begin{array}{c}
  \pim\alpha \\
 i\gm \pipl\partial_m \alpha
\end{array}\right)\right\rfloor_{r=0}
\label{cc}\eeq has an unique solution.

Now, the solutions to (\ref{ed}) and (\ref{ci}) are \beq
\Psi(x_m,\omega,\lambda)=\Psi_0 \exp({-\Lambda x_m})\, ,\nn\eeq
where $\Lambda=+\sqrt{\omega^2-\lambda}$. Note that $\Re
(\Lambda)> 0$ for $\lambda \in \mathbf{C}-\mathbf{R}_+$.

The condition (\ref{cc}), when applied to them, reads \beq
\left(\begin{array}{cc}
  \pim& 0 \\
  -\ga \omega_a  \pimi& i \gm  \pipl
\end{array}\right) \left(\begin{array}{c}
  \Psi_0 \\
  -\Lambda \Psi_0
\end{array}\right)=\left(\begin{array}{c}
  \pim\alpha \\
 -i\gm \pipl\Lambda\alpha
\end{array}\right)\,.\nn
\eeq This gives a system of two equations. After  multiplying the
second one by $i\gm$, and using $\Psi_0= \pim \Psi_0 + \pip
\Psi_0$, one obtains \beq \pim \Psi_0 =\pim \alpha
\,,\label{e1}\eeq and
\[\left(-i\gm (\ga \omega_a)\pimi \pip +\Lambda \pipl \pip
\right)\Psi_0 =\]\beq \Lambda\pipl \alpha -\left(-i\gm (\ga
\omega_a)\pimi \pim +\Lambda\pipl \pim
\right)\Psi_0\,.\label{e2}\eeq We use (\ref{prod}) and substitute
(\ref{e1}) into (\ref{e2}) to see
\[
\exp{(-\Th \gf)} \left[-i\gm (\ga \omega_a)\sinh{(\Th \gf)}+\La
\cosh{(\Th \gf)} \right]\pip \Psi_0=
\]
\beq \La \pipl \alpha - \left[-i\gm (\ga \omega_a)\cosh{(\Th
\gf)}+\La \sinh{(\Th \gf)} \right]\exp{(-\Th \gf)}\pim
\alpha\,.\nn\eeq

\bigskip

So, the problem has an unique solution if the matrix

\beq M= -i\gm (\ga \omega_a)\sinh{(\Th \gf)}+\La \cosh{(\Th \gf)}
=A \sinh{(\Th)}+\La \cosh{(\Th \gf)}\nn\eeq is nonsingular, where
we introduced $A=-i\tilde\gamma \gamma_m \gamma_a \omega_a$.

But $A^{\star}=A$; so, it is diagonalizable. Moreover,
$A^2=\left(\sum_a \omega_a ^2\right)\mathbf{1}$. As a consequence,
$A$ has eigenvalues $\pm \sqrt{\sum_a \omega_a ^2}$, except in two
dimensions, where A is proportional to the identity. Then, the
determinant of M can be evaluated in the basis of eigenvectors of
$A$, and in all cases,
$det M$ can be seen to vanish if $\lambda=\frac{\sum_a \omega_a
^2}{\cosh^2{\Th}}$. For $\lambda =0$, the determinant can only
vanish if $\omega_a =0$. Otherwise, it can only happen for
$\lambda \in \mathbf{R}_+$, which proves strong ellipticity in
$\mathbf{C}-\mathbf{R}_+$, and any even dimension. This completes
the proof of Theorem \ref{sell}. \qedbox

\section{Heat kernel in an infinite cylinder}
\label{Sect4}

In what follows, we present the heat kernel for $(P^2,{\cal B})$
in an infinite cylinder ${\cal M}= \reals _+ \times {\cal N}$ of
any even dimension. By cylinder we mean that the metric is of the
type $ds^2=dx_m^2 + ds_{{\cal N}}^2$,  where $ds_{{\cal N}}^2$ is
the metric of the closed boundary ${\cal N}$.

In order to determine the heat kernel, it is useful to note that
the chiral bag boundary conditions in equations (\ref{bc1}) and
(\ref{bc2}) are equivalent, for each eigenvalue of the tangential
part $B$ of the operator $P$, to Dirichlet boundary conditions on
part of the fibre, and Robin (modified Neumann) on the rest.

In fact, let's first notice that the operators ${\cal
P}_+=\frac{\pip \pipl}{\cosh^2{\Th}}$ and ${\cal P}_-=\frac{\pimi
\pim}{\cosh^2{\Th}}$ are self-adjoint projectors, and they satisfy
${\cal P}_++{\cal P}_-=\textbf{1}$ splitting $V$ into two
complementary subspaces.

Let $\xi=x_m -x_m^{'}$, and $\eta=x_m+x_m^{'}$ and as before, let
$y=(y_1, y_2, ...,y_{m-1})$ be the coordinates on the boundary and
$x=(y, x_m)$. If we call $\phi_{\omega}(y)$ the eigenspinors of
the operator $B=\gf \gm \gamma_a \partial_a$ (with $a= 1, 2,...,
m-1$) corresponding to the eigenvalue $\omega$, normalized such
that \beq \sum_{\omega} \phi_{\omega}^{\star}(y)
\phi_{\omega}(y^{'})= \delta^{m-1}(y-y^{'})\nn\eeq with
$\delta^{m-1}$ the Dirac delta function, and \beq
\int\limits_{\partial M}\,\,dy\,\, \phi_{\omega}^{\star}(y)
\phi_{\omega}(y)=1\,,\nn\eeq we can expand
$\psi(y,x_m)=\sum_{\omega}f_{\omega}(x_m) \phi_{\omega}(y)$. If
$\psi={\cal P}_+ \psi$, then the condition (\ref{bc1}) is
identically satisfied, and only (\ref{bc3}) must be imposed at the
boundary which, for each $\omega$, reduces to
\[\cosh{\Th}e^{-\Th \gt}\left(\partial_m +\omega \tanh{\Th}\right)f_{\omega}=0\,.\]
Since the factor to the left of the parenthesis is invertible,
this is nothing but a Robin boundary condition.

In the subspace $\psi={\cal P}_- \psi$, the boundary condition
(\ref{bc1}) reduces to\[\cosh{\Th}e^{\Th \gt}f_{\omega}=0\,,\]
while (\ref{bc3}) requires \[\omega f_{\omega}=0\,.\] Thus, in
this subspace, both boundary conditions are nothing but
homogeneous Dirichlet ones.

As a consequence, the complete heat kernel can be written as a
Dirichlet heat kernel on ${\cal P}_-V$ and a Robin heat kernel on
${\cal P}_+ V$. For the convenience of the reader we make the
single ingredients explicit \cite{cars86b} and write \beq
K(t;x,x') &=& K (t;x,x') ( {\cal P} _- +
{\cal P} _+) \nn\\
&=& \frac 1 {\sqrt{4\pi
t}}\sum_{\omega}\phi_{\omega}^{\star}(y^{'}) \phi_{\omega}(y)
e^{-\omega^2
t}\left(e^{\frac{-\xi^2}{4t}}-e^{\frac{-\eta^2}{4t}}\right){\cal
P}_- \nn\\
&+&\frac 1 {\sqrt{4\pi
t}}\sum_{\omega}\phi_{\omega}^{\star}(y^{'}) \phi_{\omega}(y)
e^{-\omega^2
t}\left\{\left(e^{\frac{-\xi^2}{4t}}+e^{\frac{-\eta^2}{4t}}\right)\right.
\nn\\
& &\left.+2 \sqrt{\pi t}\,\, \omega \tanh \theta e^{\omega ^2 t
\tanh ^2 \theta - \omega \eta\tanh \theta } erfc [ u_\omega (\eta
, t)]
\right\} {\cal P}_+ \nn\\
&=&\frac{1}{\sqrt{4\pi
t}}\sum_{\omega}\phi_{\omega}^{\star}(y^{'}) \phi_{\omega}(y)
e^{-\omega^2
t}\left\{\left(e^{\frac{-\xi^2}{4t}}-e^{\frac{-\eta^2}{4t}}\right)\mathbf{1}\right.\label{hk}\\
& & \left.+\frac{2\pip \pipl}{\cosh^2(\Th)}\left[1+\sqrt{(\pi
t)}\omega \tanh{\Th}e^{u_{\omega}(\eta,t)^2}
erfc[u_{\omega}(\eta,t)]\right]e^{\frac{-\eta^2}{4t}}\right\}
\nn\eeq where
$u_{\omega}(\eta,t)=\frac{\eta}{\sqrt{4t}}-\sqrt{t}\omega\tanh(\Th)$,
and\[erfc(x)=\frac{2}{\sqrt{\pi}}\int_{x}^{\infty}d\xi
e^{-\xi^2}\,\] is the complementary error function. Note that
(\ref{hk}) is a direct generalization of the heat kernel given in
\cite{bene02-35-9343} for the two-dimensional case, which, in
turn, coincides with the Fourier transform of equation (101) in
\cite{duer97-255-333} for an antiperiodic boundary fiber.

\section{Meromorphic properties of the zeta function}
\label{Sect6}

Let us now analyze the boundary contributions to the global zeta
function related to (\ref{hk}). We first note that global
quantities are necessarily divergent due to the non-compact nature
of our manifold ${\cal M} = \reals _+ \times {\cal N}$. This is
not a severe problem because the result (\ref{hk}) allows us to
identify easily the bulk term leading to a divergent contribution
when integrated. In particular, it is the first term in (\ref{hk})
that represents the heat kernel on the manifold $\reals \times
{\cal N}$. In the following, without changing the notation, we
will first ignore this term and this will allow us to determine
the boundary contributions to the global zeta function.
Alternatively, as we will show afterwards, one can introduce a
localizing function of compact support in (\ref{hk}) such that the
trace gives a finite result.

Let us consider the trace of ({\ref{hk}) ignoring the first term.
It is convenient to perform the Dirac trace ($tr$) first. Since
\[tr \left(\frac{2\pip \pipl}{\cosh^2(\Th)}\right)=2^{\bar
m}\,,\] the trace of the 'boundary' heat kernel reduces to \beq Tr
K(t;x,x)&=& \frac{2^{\bar m}}{2}\sum_{\omega} \omega
\tanh{\Th}e^{-\omega^2 t}\nn\\
& &\times \,\,\int_0^{\infty} dx_m\, erfc[u_{\omega}(2x_m ,
t)]e^{\frac{-x_m^2}{t}+ u_\omega ^2 (2x_m ,\,\, t) }\,,\nn\eeq
where the second and third term in (\ref{hk}) have cancelled each
other. Now, using that
\[ -\frac12 \frac{\partial}{\partial x_m}
\left[e^{-x_m^2/t+u_{\omega}^2(2x_m,t)}
erfc\left[u_{\omega}(2x_m,t)\right]\right] =\]\beq
e^{-x_m^2/t}\left[\frac{1}{\sqrt{\pi t}}+\omega \tanh\Th\,
e^{u_{\omega}^2(2x_m,t)}
erfc\left[u_{\omega}(2x_m,t)\right]\right]\,,\nn\eeq we get \[ Tr
K(t;x,x)= \frac{2^{\bar m}}{4}\sum_{\omega}e^{-\omega^2
t}\left[e^{u_{\omega}^2(0,t)}erfc\left[u_{\omega}(0,t)\right]-1\right]=\]\beq
\frac{2^{\bar m}}{4}\sum_{\omega}\left[e^{\frac{-\omega^2 t
}{\cosh^2{\Th}}}\left[1+erf(\omega\sqrt{t}
\tanh{\Th})\right]-e^{-\omega^2 t}\right]\,.\nn\eeq Here we used
$erf(x)=-erf(-x)=1-erfc(x)$.

Now, we can Mellin transform this trace, to obtain the 'boundary'
zeta function of the square of the Dirac operator in the infinite
cylinder \beq \zeta(s,P^2)&=&\frac{2^{\bar
m}}{4\Gamma(s)}\sum_{\omega}\int_0^{\infty}dt \,\,
t^{s-1}\left[e^{\frac{-\omega^2
t }{\cosh^2{\Th}}}-e^{-\omega^2 t}\right]\nn\\
& &+ \frac{2^{\bar
m}}{4\Gamma(s)}\sum_{\omega}\int_0^{\infty}dt\,\,t^{s-1}e^{\frac{-\omega^2
t }{\cosh^2{\Th}}}erf(\omega\sqrt{t}
\tanh{\Th})\nn\\
&=&\zeta_1(s,P^2)+\zeta_2(s,P^2)\,.\label{z}\eeq

The first contribution can be readily seen to be \beq
\zeta_1(s,P^2)=\frac{1}{4}\left(\cosh^{2s}
{\Th}-1\right)\zeta(s,B^2)\,,\label{z1}\eeq where $B$ is the
operator defined in Section \ref{Sect4}.

As for the second contribution to (\ref{z}), it is given by \beq
\zeta_2(s,P^2)=\frac{2^{\bar
m}}{4\Gamma(s)}\sum_{\omega}\int_0^{\infty}dt\,\,t^{s-1}e^{\frac{-\omega^2
t }{\cosh^2{\Th}}}\frac{2}{\sqrt{\pi}}\int_0^{(\omega\sqrt{t}
\tanh{\Th})} d\xi e^{-\xi^2}\,.\nn\eeq

After changing variables according to
$y=\frac{\xi\cosh\Th}{\sqrt{t}\omega}$, and interchanging
integrals, one finally gets \beq \zeta_2(s,P^2)&=&\frac{2^{\bar
m}\Gamma\left(s+\frac{1}{2}\right)}{4\Gamma(s)}\cosh^{2s} \Th
\sum_{\omega}sign(\omega) \left(\omega^2\right)^{-s}\nn\\
& &\quad \quad \quad \quad \quad \quad \times \,\,
\frac{2}{\sqrt{\pi}}\int_0^{\sinh{\Th}} dy
\left(1+y^2\right)^{-s-\frac{1}{2}}\nn\\
&=& \frac{\Gamma\left(s+\frac{1}{2}\right)}{4\Gamma(s)}\cosh^{2s}
\Th\eta(2s,B) \frac{2}{\sqrt{\pi}}\int_0^{\sinh{\Th}}
dy\left(1+y^2\right)^{-s-\frac{1}{2}}\nn\\
&=&\frac 1 {2 \sqrt{\pi}} \frac{\Gamma\left( s + \frac  1 2
\right) } {\Gamma (s)} \sinh \theta \cosh ^{2s} \theta \eta (2s,
B)\nn\\
& &\quad \times \,\,_2F_1 \left( \frac 1 2 , \frac 1 2 +s, \frac 3
2 ; -\sinh ^2 \theta \right)\,. \label{z2}\eeq The structure of
the zeta function is similar to the structure found for spectral
boundary conditions, see e.g. \cite{grub96-6-31}. In particular,
the analysis of the zeta function on ${\cal M}$ has been reduced
to the analysis of the zeta and eta function on ${\cal N}$.

As already commented, from (\ref{z1}) and (\ref{z2}) one can
determine the positions of the poles and corresponding residues
for the zeta function in any cylindrical product manifold, in
terms of the meromorphic structure of the zeta and eta functions
of the boundary operator. For the rightmost poles, explicit
results can be given in terms of the geometry of the boundary
${\cal N}$. For example, for $s=(m-1)/2$ we see that \beq
\mbox{Res } \zeta_1 \left( \frac {m-1} 2 , P^2 \right) &=& \frac 1
4 \left( \cosh ^{m-1} \theta -1 \right) \mbox{Res } \zeta \left(
\frac {m-1} 2 , B^2 \right) \nn\\
&=& \frac 1 4 \left( \cosh ^{m-1} \theta -1 \right) \frac{ (4\pi )
^{-(m-1)/2} } {\Gamma \left( \frac{m-1} 2 \right)}\,\, 2^{\bar
m}\,\,\mbox{Vol} ({\cal N}).\nn\eeq Because $\zeta_2$ does not
contribute, given $\eta (2s,B)$ is regular at $s=(m-1)/2$
\cite{gilk95b}, this equals $\mbox{Res }\zeta (s,P)$ and is the
result expected from the calculation on the ball
\cite{kirs02-104-119}. For $\theta =0$ the residue disappears as
is known to happen for the standard local bag boundary conditions
\cite{gilk95b}. Further results can be obtained by using Theorem
4.4.1 of \cite{gilk95b}. Given we considered the case without
'potential', it is immediate that \beq \mbox{Res } \zeta \left(
\frac{ m-2} 2 , P^2 \right) =0.\nn\eeq Also, for the particular
case of $s=0$, the fact that $\zeta (s,B^2)$ and $\eta (2s,B)$ are
regular at $s=0$ shows that $\zeta (0,P^2) =\zeta(0,P)=0$.

Given the local heat kernel (\ref{hk}), a local version of the
results of this section is easily obtained. To this end, we use a
localizing function with compact support near the boundary, such
that its normal derivatives at the boundary vanish, \beq
\left.\frac{\partial ^n} {\partial x_m^n} f(y,x_m) \right|_{x_m
=0} =0 , \quad n\in\nats.\nn \eeq Furthermore, we let $\tilde P^2$
denote the operator $P^2$ on the double $\mathbf{R} \times {\cal
N}$ of $\mathbf{R}_+\times {\cal N}$, and we extend the localizing
function as an even function to the double. We use the notation
$f=f(y,x_m)$, $f_{{\cal N}} = f(y,x_m =0)$ and $\tilde f$ for $f$
on the double. Introducing the local versions $\zeta (\tilde f, s,
\tilde P^2)$, $\zeta (f,s,P^2)$, $\zeta (f_{{\cal N}}, s, B^2)$
and $\eta (f_{{\cal N}},2s, B)$ of the zeta functions and the eta
function, (\ref{hk}) and previous calculations show that the
following theorem holds:
\newtheorem{local}{Theorem}[section]
\begin{local}
\beq \lefteqn{ \Gamma (s) \zeta (f,s,P^2) = \Gamma (s) \left\{
\frac 1 2 \zeta
(\tilde f , s , \tilde P ^2) \right. } \nn\\
 & &+ \frac 1 4 \left( \cosh ^{2s} \theta -1 \right) \zeta
 (f_{{\cal N}}, s, B^2 ) \label{th2}\\
 & &\left. + \frac 1 {2\sqrt{\pi}} \frac{\Gamma (s+1/2)} {\Gamma (s)}
 \sinh \theta \cosh ^{2s} \theta \phantom{a}_2F_1 \left( \frac 1 2
 , \frac 1 2 +s, \frac 3 2 , -\sinh ^2 \theta \right) \eta
 (f_{{\cal N}}, 2s, B ) \right\}\nn\\
 & &+ h(s)\nn
 \eeq
 where $h(s)$ is an entire function.
\end{local}
This result parallels Theorem 2.1 in \cite{grub96-6-31} for
spectral boundary conditions.
\section{Conclusions}
In this article we have considered the influence of generalized
bag boundary conditions on the heat kernel and the zeta function.
In order to guarantee certain structural properties we have first
shown the strong ellipticity of the boundary conditions. Work by
Seeley \cite{seel68-10-288,seel69-91-963} then shows the standard
heat kernel expansion holds and so the zeta function can have only
simple poles at $s=m/2,(m-1)/2,...,1/2$, and $s=-(2l+1)/2$,
$l\in\nats$. This is the main result of this paper.

Based on the strong ellipticity one might envisage the
determination of the leading heat kernel coefficients for
generalized bag boundary conditions as they are needed for the
calculation of effective actions in gauge theories in Euclidean
bags \cite{wipf95-443-201}. Special case calculations can serve to
restrict the general form that coefficients can have, cylindrical
manifolds providing a valuable example. Here, for
$P=i\gamma_j\nabla_j$, we have expressed the heat kernel and the
zeta function of the associated second order operator on ${\cal M}
= \reals_+ \times {\cal N}$ in terms of the boundary data on
${\cal N}$. In fact, this result, under certain restrictions, can
be straightforwardly generalized to $P=i\gamma_j\nabla_j - \phi$.
In order that a separation of variables as presented succeeds we
need $\partial _{x_m} \phi =0$ and $\{\gamma_m , \phi \} = \{\gt ,
\phi \} =0$. If this is satisfied, equations (\ref{z1}) and
(\ref{z2}) remain valid, once the operator $B$ incorporates the
potential, $B = \gt \gamma_m (\gamma_a \nabla_a - i\phi )$. We
have thus a particular case involving a potential and Riemannian
curvature and various restrictions on heat kernel coefficients
will follow.

\renewcommand{\theequation}{\Alph{section}.\arabic{equation}}
\begin{appendix}

\section{Appendix: $\gamma$-matrices}
Let $m=2\bar m$ be the dimension of a Riemannian manifold. We
denote by $\gamma_j^{(m)}$, $j=1,...,m,$ the self-adjoint $\gamma$-matrices
projected along some $m$-bein system. These are defined inductively by
\beq
\gamma_j^{(m)} = \left(
\begin{array}{cc}
0 & i\gamma_j ^{(m-2)} \\
-i \gamma_j ^{(m-2)} & 0
\end{array}
\right) , & & j=1,...,m-1,\nn\\
\gamma_m^{(m)} = \left(
\begin{array}{cc}
0 & 1 \\
1 & 0
\end{array}
\right) , & & \gamma^{(m)}_{m+1} = \left(
\begin{array}{cc}
1 & 0 \\
0 & -1
\end{array}
\right),\nn \eeq starting from the Pauli matrices \beq \gamma_1
^{(2)} = \left( \begin{array}{cc}
0 & i \\
-i & 0
\end{array}
\right)  , \quad
\gamma_2 ^{(2)} = \left( \begin{array}{cc}
0 & 1 \\
1 & 0
\end{array}
\right)  , \quad \gamma_3^{(2)} = \left( \begin{array}{cc}
1 & 0 \\
0 & -1
\end{array} \right) . \nn
\eeq The $\gamma$-matrices satisfy the Clifford anti-commutation
formula \beq \gamma_j ^{(m)} \gamma_k ^{(m)} + \gamma_k^{(m)}
\gamma_l ^{(m)} = 2 \delta _{kl} .\label{A3} \eeq In the main body
of the paper we will simplify the notation and we will not
indicate the dimension explicitly. In addition, we set \beq
\gamma_{m+1}^{(m)} = \gt = (-i)^{\bar m} \gamma_1 ... \gamma_m,
\nn\eeq which is the generalization of `$\gamma_5$' to arbitrary
even dimension.

\end{appendix}
$\phantom{aa}$\\
{\bf Acknowledgements:} Research of CGB and EMS was partially
supported by CONICET(PIP 0459/98) and UNLP(11/X298). Research of
PG was partially supported by the MPI (Leipzig, Germany). KK
acknowledges support by the Baylor University Summer Sabbatical
Program and by the MPI (Leipzig, Germany).


\begin{thebibliography}{10}


\bibitem{kirs01}
K.~Kirsten.
\newblock {\em Spectral Functions in Mathematics and Physics}.
\newblock Chapman\&Hall/CRC, Boca Raton, FL, 2001.

\bibitem{hras84-245-118}
P.~Hrasko and J.~Balog.
\newblock The fermion boundary condition and the theta angle in {Q}{E}{D} in
  two-dimensions.
\newblock {\em Nucl. Phys.}, B245:118--126, 1984.

\bibitem{wipf95-443-201}
A.~Wipf and S.~D{\"u}rr.
\newblock Gauge theories in a bag.
\newblock {\em Nucl. Phys.}, B443:201--232, 1995.

\bibitem{duer97-255-333}
S.~D{\"u}rr and A. Wipf. \newblock Finite temperature Schwinger
model with chirality breaking boundary conditions.
\newblock {\em Ann. Phys.}, 255:333--361, 1997.

\bibitem{duer99-273-1}
S.~D{\"u}rr. \newblock Aspects of quasi-phasestructure of the
Schwinger model on a cylinder with broken chiral symmetry.
\newblock {\em Ann. Phys.}, 273:1-36, 1999.

\bibitem{bene02-35-9343}
C.G. Beneventano, E.M. Santangelo, and A.~Wipf.
\newblock Spectral asymmetry for bag boundary conditions.
\newblock {\em J. Phys. A: Math. Gen.}, 35:9343--9354, 2002.

\bibitem{espo02-66-085014}
G.~Esposito and K.~Kirsten.
\newblock Chiral bag boundary conditions on the ball.
\newblock {\em Phys. Rev.}, D66:085014, 2002.

\bibitem{dowk99-16-1917}
J.S. Dowker and K.~Kirsten.
\newblock The a(3/2) heat kernel coefficient for oblique boundary conditions.
\newblock {\em Class. Quantum Grav.}, 16:1917--1936, 1999.

\bibitem{avra99-200-495}
I.G. Avramidi and G.~Esposito.
\newblock Gauge theories on manifolds with boundary.
\newblock {\em Commun. Math. Phys.}, 200:495--543, 1999.

\bibitem{grub96-6-31}
G.~Grubb and R.T. Seeley.
\newblock Zeta and eta functions for {A}tiyah-{P}atodi-{S}inger operators.
\newblock {\em J. Geom. Anal.}, 6:31--77, 1996.

\bibitem{1}
E. Elizalde and D.V. Vassilevich.
\newblock Heat kernel coefficients for Chern-Simons boundary
conditions in QED.
\newblock {\em Class. Quantum Grav.}, 16:813--823, 1999.

\bibitem{2a}
E. Elizalde, M. Lygren and D.V. Vassilevich.
\newblock Antisymmetric tensor fields on spheres: Functional
determinants and nonlocal counterterms.
\newblock {\em J. Math. Phys.}, 37:3105--3117, 1996.

\bibitem{2b}
E. Elizalde, M. Lygren and D.V. Vassilevich.
\newblock Zeta function for the Laplace operator acting on forms
in a ball with gauge boundary conditions.
\newblock {\em Commun. Math. Phys.}, 183:645--660, 1997.

\bibitem{3}
J. {\'S}niatycki and G. Schwarz. \newblock The existence and
uniqueness of solutions of Yang-Mills equations with bag boundary
conditions. \newblock {\em Comm. Math. Phys.}, 159:593--604, 1994.

\bibitem{5}
B.H.J. McKellar, G.J. Stephenson, Jr., and M.J. Thomson. \newblock
The Dirac equation in Kerr spacetime, spheroidal coordinates and
the MIT bag model of hadrons. \newblock {\em J. Phys. A: Math.
Gen.}, 26:3649--3657, 1993.

\bibitem{6}
B. Freedman and V. Krapchev. \newblock  Effects of boundary
conditions on massless two-dimensional electrodynamics in a static
bag. \newblock {\em Phys. Rev. D}, 14:566--577, 1976.

\bibitem{7}
D.A. Geffen and H. Suura. \newblock Solutions to a
gauge-invariant, equal-time two-body wave equation. Light-mass
quark-antiquark system. \newblock {\em Phys. Rev. D},
16:3305--3319, 1977.

\bibitem{8}
K. Colanero and M.-C. Chu. \newblock Analytical solution of the
dynamical spherical MIT bag. \newblock {J. Phys. A: Math. Gen.},
35:993--999, 2002.

\bibitem{9}
S. Ansoldi, C. Castro and E. Spallucci. \newblock Chern-Simons
hadronic bag from quenched large-$N$ QCD. \newblock {\em Phys.
Lett. B}, 504:174--180, 2001.

\bibitem{12}
V.V. Kravchenko. \newblock On a biquaternionic bag model.
\newblock {\em Z. Anal. Anwend.}, 14:3--14, 1995.

\bibitem{gilk95b}
P.B. Gilkey.
\newblock {\em Invariance Theory, the Heat Equation and the Atiyah-Singer Index
  Theorem}.
\newblock CRC Press, Boca Raton, 1995.

\bibitem{cars86b}
H.S. Carslaw and J.C. Jaeger
\newblock {\em Conduction of Heat in Solids}.
\newblock Clarendon Press, Oxford, 1986.

\bibitem{kirs02-104-119}
K.~Kirsten.
\newblock Heat kernel asymptotics: more special case calculations.
\newblock {\em Nucl. Phys. B (Proc. Suppl.)}, 104:119--126, 2002.

\bibitem{seel68-10-288}
R.T. Seeley.
\newblock Complex powers of an elliptic operator, {S}ingular {I}ntegrals,
  {C}hicago 1966.
\newblock {\em Proc. Sympos. Pure Math.}, 10:288--307, American Mathematics
  Society, Providence, RI, 1968.

\bibitem{seel69-91-963}
R.T. Seeley.
\newblock Analytic extension of the trace associated with elliptic boundary
  problems.
\newblock {\em Amer. J. Math.}, 91:963--983, 1969.


\end{thebibliography}

\end{document}